\documentstyle[11pt,epsfig]{article}

\begin{document}

\title{\bf Checking the Burkardt sum rule for the Sivers function
by model calculations}

\author{K.~Goeke, S.~Mei{\ss}ner, A.~Metz, and M.~Schlegel
 \\[0.3cm]
{\it Institut f\"ur Theoretische Physik II,} \\
{\it Ruhr-Universit\"at Bochum, D-44780 Bochum, Germany}}

\date{\today}
\maketitle

\begin{abstract} 
\noindent
The Sivers mechanism gives rise to a non-zero average transverse 
momentum of partons inside a transversely polarized target.
According to a sum rule of Burkardt this transverse momentum vanishes
when summing over all partons. 
We explicitly check the Burkardt sum rule by means of one-loop 
calculations for a quark target in perturbative QCD and for
a diquark spectator model of the nucleon.
In both cases the sum rule is fulfilled.
\end{abstract}

\noindent
{\bf 1.} In the case of a transversely polarized target with 4-momentum 
$P$ and spin vector $S$ one can have a correlation of the type 
$\vec{S}_T \cdot (\vec{P} \times \vec{k}_T)$, where 
$\vec{k}_T$ is the transverse momentum of an arbitrary parton inside 
the target.
The strength of this correlation is quantified in terms of the Sivers
parton distribution~\cite{sivers_89,sivers_90} (called $f_{1T}^{\perp}$ 
in the notation of Refs.~\cite{boer_97c,mulders_95}).
The discovery that the Sivers effect is not forbidden by time-reversal
invariance of the strong interaction~\cite{brodsky_02a,collins_02} 
initiated an enormous amount of subsequent work.
The Sivers effect is of particular interest because it may be at the
origin of various observed single spin asymmetries in hard scattering
processes.

In the meantime not only model 
calculations~\cite{brodsky_02a,collins_02,ji_02,gamberg_03a,yuan_03a,bacchetta_03b}
of the Sivers function are available but also experimental data 
already exists.
The Sivers asymmetry has recently been measured in semi-inclusive 
deep-inelastic scattering (DIS) by the 
HERMES~\cite{airapetian_04,diefenthaler_05} and 
COMPASS~\cite{alexakhin_05} Collaborations, where in particular in the 
case of a proton target a non-zero effect has been 
observed~\cite{airapetian_04,diefenthaler_05}.
Using the available data already first extractions of the Sivers function 
have been 
performed~\cite{efremov_04a,anselmino_05a,anselmino_05b,vogelsang_05,collins_05a} 
(see Ref.~\cite{anselmino_05d} for a comparison of different fits).
On the basis of these fits also predictions for the Sivers asymmetry
appearing in other processes like the Drell-Yan reaction have been 
made~\cite{efremov_04a,anselmino_05a,anselmino_05b,vogelsang_05,collins_05b}.

In this Letter we are concerned with an interesting sum rule for the
Sivers function, that was derived by Burkardt for a spin-$\frac{1}{2}$
target in QCD~\cite{burkardt_04,burkardt_03}.
This sum rule states that the average transverse momentum induced 
by the Sivers effect vanishes if one sums over all 
partons~\cite{burkardt_04},
\begin{equation} \label{e:sum_1}
\sum_{a = q,\bar{q},g} \langle \vec{k}_T^a \rangle = 0 \,.
\end{equation}
Expressed in terms of the Sivers function it takes the 
form~\cite{efremov_04a,burkardt_04}
\begin{equation} \label{e:sum_2}
\sum_{a = q,\bar{q},g} \int_0^1 dx \, f_{1T}^{\perp(1)a}(x) = 0 \,,
 \quad \textrm{with} \quad
f_{1T}^{\perp(1)a}(x) 
 = \int d^2\vec{k}_T \frac{\vec{k}_T^2}{2M^2} f_{1T}^{\perp a}(x,\vec{k}_T^2) 
\end{equation}
and $M$ denoting the mass of the target.
The Burkardt sum rule can serve as a non-trivial constraint when
modelling or fitting the Sivers function.
Moreover, the sum rule in combination with large-$N_c$ 
arguments~\cite{pobylitsa_03} implies that the Sivers function for 
the gluon, which has been considered in various recent 
articles~\cite{boer_03b,anselmino_04a,schmidt_05,anselmino_05c},
is suppressed in the limit of a large number of colors~\cite{efremov_04a}.

Because of its fundamental character it is worthwhile to explicitly 
check the Burkardt sum rule by model calculations.
To this end we compute the Sivers function of a quark and a gluon for 
a quark target in perturbative QCD. 
The calculation is limited to the one-loop approximation.
We also investigate the sum rule in the framework of a simple diquark
spectator model of the nucleon~\cite{brodsky_02a}.
We find that in both cases the sum rule is fulfilled.
As a by-product our study provides the first calculation of the Sivers 
function for a gluon and a scalar particle.
\\[0.3cm]
\noindent
{\bf 2.}~We begin by specifying our conventions.
The $k_T$-dependent distribution of quarks in a transversely polarized 
spin-$\frac{1}{2}$ target is given by~\cite{boer_97c,bacchetta_04b}
\begin{eqnarray} \label{e:quark}
\Phi^{q}(x,\vec{k}_{T},S) & = & 
f_{1}^{q}(x,\vec{k}_{T}^{2}) 
- \frac{\varepsilon_{T}^{ij} k_{Ti} S_{Tj}}{M} \, 
f_{1T}^{\perp q}(x,\vec{k}_T^{2}) 
\\
& = & \frac{1}{2}
\int \frac{d \xi^- \, d^2 \vec{\xi}_{T}}{(2 \pi)^3} \, 
e^{i \, k \cdot \xi} \,
\langle P,S \, | \, \bar{\psi}(0) \, \gamma^+ \,
 {\cal W}(0 , \xi) \, \psi(\xi) \, | \, P,S \rangle \Big|_{\xi^+=0} \,,  
\nonumber
\end{eqnarray}
with $\varepsilon_T^{ij} \equiv \varepsilon^{-+ij}$ ($\varepsilon^{0123} = 1$)
and $k_T^{\mu} = (0,0,\vec{k}_T)$.
While the $k_T$-dependent unpolarized quark distribution $f_1^q$ is an even
function of $\vec{k}_T$, the Sivers effect is odd and thus leads to an 
azimuthal asymmetry of $\Phi^q$ about the direction of the target momentum. 
The Wilson line ${\cal W}(0 , \xi)$ ensures color gauge invariance of the 
bilocal quark operator, and its presence is crucial for otherwise the Sivers
effect would vanish when applying time-reversal 
to~(\ref{e:quark})~\cite{collins_92b,collins_02}.
The Wilson line is defined
through~\cite{collins_81c,collins_02,ji_02,belitsky_02,boer_03a}  
\begin{equation} \label{e:path} 
 {\cal W}(0, \xi) =
 [0,0,\vec{0}_{T} ; 0,\infty,\vec{0}_{T}]
 \mbox{} \times
 [0,\infty,\vec{0}_{T};0,\infty,\vec{\xi}_{T}]
 \mbox{} \times
 [0,\infty,\vec{\xi}_{T};0,\xi^-,\vec{\xi}_{T}] \,,
\end{equation}
where $[a^+,a^-,\vec{a}_{T};b^+,b^-,\vec{b}_{T}]$ denotes a gauge link
connecting the points $a^\mu=(a^+,a^-,\vec{a}_{T})$ and
$b^\mu=(b^+,b^-,\vec{b}_{T})$ along a straight line.
To be specific we have for instance
\begin{equation} \label{e:link}
[0,0,\vec{0}_{T} ; 0,\infty,\vec{0}_{T}]
= {\cal P} \exp \bigg( -i g \int_0^{\infty} dz^- A^+(0,z^-,\vec{0}_T) \bigg) \,,
\end{equation}
where the overall sign in the exponent is determined by the covariant 
derivative $D^{\mu} = \partial^{\mu} - i g A^{\mu}$.
The gluon field in (\ref{e:link}) is in the fundamental representation, i.e.,
$A^{\mu} = A_a^{\mu} \lambda_a / 2$ with $\lambda_a$ representing the Gell-Mann
matrices.
The gauge link in Eq.~(\ref{e:path}) with future-pointing Wilson lines is 
suitable for semi-inclusive DIS, whereas past-pointing lines are needed 
for the Drell-Yan process.
The different paths of the Wilson lines lead to a different sign of the 
Sivers function in both reactions~\cite{collins_02}.
The results we present below are obtained using the link in~(\ref{e:path}).
Eventually, we note that by definition only connected diagrams are considered
when evaluating the correlator~(\ref{e:quark}), and that for a colored target 
(like the quark target we are going to use) an average over the color of the 
target is implicit.

Analogous to Eq.~(\ref{e:quark}) the distributions of antiquarks and gluons
read~\cite{collins_81c,burkardt_04}
\begin{eqnarray} \label{e:antiquark}
\Phi^{\bar{q}}(x,\vec{k}_{T},S) & = & \frac{1}{2}
\int \frac{d \xi^- \, d^2 \vec{\xi}_{T}}{(2 \pi)^3} \, 
e^{i \, k \cdot \xi} \,
\langle P,S \, | \, \textrm{Tr} \Big[ \gamma^+ \psi(0) \,
 {\cal W}^{\ast}(0 , \xi) \, \bar{\psi}(\xi) \Big] \, 
 | \, P,S \rangle \Big|_{\xi^+=0} \,,
\\ \label{e:gluon}
\Phi^{g}(x,\vec{k}_{T},S) & = & \frac{1}{x P^+}
\int \frac{d \xi^- \, d^2 \vec{\xi}_{T}}{(2 \pi)^3} \, 
e^{i \, k \cdot \xi} \,
\langle P,S \, | \, F_a^{+i}(0) \, {\cal W}_{ab}(0 , \xi) 
\, F_b^{+i}(\xi) \, | \, P,S \rangle \Big|_{\xi^+=0} \,.
\end{eqnarray}
In~(\ref{e:gluon}) the field strength tensor $F^{\mu\nu}$ enters and the
gluon field in the Wilson line is in the adjoint representation 
($A_{ab}^{\mu} \equiv - i f_{abc} A_c^{\mu}$ with $f_{abc}$ denoting the 
structure constant of the SU(3) group).
Our definition of the gluon parton density differs from that of 
Ref.~\cite{burkardt_04} by a factor of two.
When evaluating the sum rule~(\ref{e:sum_1}, \ref{e:sum_2}) this difference gets 
compensated because, in contrast to Ref.~\cite{burkardt_04}, we consider the 
gluon distribution only for $x > 0$.
Expressed in terms of the distributions $\Phi^a$ 
in~(\ref{e:quark}, \ref{e:antiquark}, \ref{e:gluon}) the Burkardt sum rule 
reads~\cite{burkardt_04}
\begin{equation} \label{e:rule}
\sum_{a = q,\bar{q},g} \langle \vec{k}_T^a \rangle \equiv
\sum_{a = q,\bar{q},g} \int_0^1 dx \int d^2\vec{k}_T \, \vec{k}_T \, 
\Phi^a(x,\vec{k}_{T},S) = 0 \,.
\end{equation}
Although it is obvious that the unpolarized parton distribution $f_1$
does not contribute to the sum rule, the cancellation of the average
transverse momentum due to the Sivers effect (when summing over all
partons) is not immediately evident.
The non-trivial character of the sum rule also becomes clear from the 
derivation which, in particular, makes use of the light-cone gauge and 
the finiteness of the light-cone energy 
at $\xi^- = \pm \infty$~\cite{burkardt_04,burkardt_03}.
In principle the Burkardt sum rule can be considered to be as fundamental 
as the famous longitudinal momentum sum rule for the unpolarized forward 
parton density~\cite{collins_81c}.
Two issues, however, may require additional investigation:
first, the treatment of UV-divergences which appear when integrating upon 
transverse parton momenta;
second, light-cone divergences which can be caused by the light-like Wilson 
lines in~(\ref{e:path})~\cite{collins_81c,collins_03}.
These light-cone divergences don't show up in the simple one-loop 
calculations of the Sivers function discussed in this Letter, but may well
appear when extending the calculations to higher orders.
In order to avoid such infinities modified definitions of $k_T$-dependent 
parton densities have been proposed and used in the literature 
(see, e.g., Refs.~\cite{collins_81c,collins_99,collins_03,ji_04a,ji_04b,collins_04}), 
where certain non-light-like Wilson lines appear.
As a consequence the $k_T$-dependent distributions also depend on the 
direction of those Wilson lines.
Though the influence of these two points should be studied, in our opinion 
it is unlikely that the validity of the Burkardt sum rule gets affected.

\begin{figure}[t]
\begin{center}
\includegraphics[width=13cm]{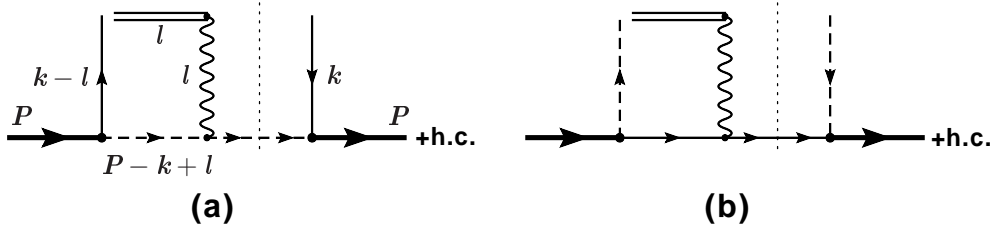}
\end{center}
\caption{One-loop diagrams relevant to the calculation of the Sivers function
for the quark (diagram (a)) and the scalar diquark (diagram (b)) in the diquark 
model.
The Hermitian conjugate diagrams (h.c.) are not shown.
We are dealing with cut diagrams indicated by the thin dashed line.
In diagram (a) for instance the diquark (dashed line) interacts through the
exchange of a photon with the eikonalized quark (double line).}
\label{f:diquark}
\end{figure}

The Burkardt sum rule was derived in QCD, but one may expect it to hold also
in different models of non-perturbative QCD.
In our study here we are interested in the scalar diquark model of the
nucleon~\cite{brodsky_02a}.
In this (Abelian) model quarks, diquarks, photons, and also the nucleon itself 
(and the respective antiparticles) appear as partons inside the nucleon, where 
to the order we are considering here only the former two are relevant.
While the quark Sivers function is computed according to the 
definition~(\ref{e:quark}), the one for the scalar diquark is obtained from
the correlator
\begin{equation} \label{e:scalar}
\Phi^{s}(x,\vec{k}_{T},S) =  x P^+
\int \frac{d \xi^- \, d^2 \vec{\xi}_{T}}{(2 \pi)^3} \, 
e^{i \, k \cdot \xi} \,
\langle P,S \, | \, \phi(0) \,
 {\cal W}(0 , \xi) \, \phi(\xi) \, | \, P,S \rangle \Big|_{\xi^+=0} \,. 
\end{equation}
The Wilson line for both quark and diquark density contains in this model the 
Abelian gauge field $A^{\mu}$, and the coupling $g$ in Eq.~(\ref{e:link}) is 
to be replaced by the negative of the electric charge $e_q$ and $e_s$ of the 
quark and scalar diquark respectively.
The sum rule we expect to hold in the diquark model is given by
\begin{equation} \label{e:rule_dq}
\sum_{a = q,s} \int_0^1 dx \int d^2\vec{k}_T \, \vec{k}_T \, 
\Phi^a(x,\vec{k}_{T},S) = 0 \,,
\end{equation}
where the sum here is understood to include also antiparticles.
Note that the Sivers function for the photon vanishes due to time-reversal 
invariance, because the definition of the $k_T$-dependent photon distribution,
which is analogous to Eq.~(\ref{e:gluon}), does not contain a gauge link.
Moreover, the Sivers function for a nucleon as a parton itself is zero in the 
diquark model of Ref.~\cite{brodsky_02a} since the nucleon is chargeless.
\\[0.3cm]
\noindent
{\bf 3.}~We now turn our attention to the results of the model calculations 
by first considering the diquark model of the nucleon~\cite{brodsky_02a}.
In this approach the interaction between the nucleon, the quark, and the 
diquark is given by a point-like scalar vertex with the coupling 
constant $\lambda$.
In order to compute the Sivers function of the quark and the diquark to 
lowest non-trivial order the diagrams in Fig.~\ref{f:diquark} have to be 
evaluated, where only the imaginary part of the loop amplitude 
(with loop-momentum $l$) contributes.
The calculation can be done by means of standard techniques, and we refrain 
from giving any details here.
One finds the results
\begin{eqnarray} \label{e:res_q1}
f_{1T}^{\perp q}(x,\vec{k}_T^2) & = & \frac{e_q e_s \lambda^2}{4(2\pi)^4}
 \frac{(1-x) M (x M + m_q)}{\vec{k}_T^2 (\vec{k}_T^2 + \tilde{m}^2)}  
 \, \ln \frac{\vec{k}_T^2 + \tilde{m}^2}{\tilde{m}^2} \,,
\\
& & \textrm{with} \quad 
\tilde{m}^2 = x(1-x) \bigg( -M^2 +\frac{m_q^2}{x} + \frac{m_s^2}{1-x} \bigg) \,,
\nonumber \\ \label{e:res_s1}
f_{1T}^{\perp s}(x,\vec{k}_T^2) & = & 
 - f_{1T}^{\perp q}(1-x,\vec{k}_T^2) \,.
 \vphantom{\frac{1}{1}}
\end{eqnarray}
While the quark Sivers function in the diquark model has already been computed 
previously~\cite{collins_02,ji_02}, the result (\ref{e:res_s1}) for the 
scalar particle is new.
Note now that for the individual partons the average transverse momentum
as defined through~(\ref{e:rule}) does not exist.
When performing the $k_T$-integration one is facing a logarithmic UV-divergence 
well known from such type of calculations.
If one regularizes this integral (by using for instance a cut-off or 
dimensional regularization) one can interchange the $x$- and $k_T$-integral 
in the evaluation of the sum rule~(\ref{e:rule_dq}).
Doing so and then replacing the integration variable for the longitudinal
momentum according to $x \to 1 - x$ for one of the partons shows that the 
sum rule~(\ref{e:rule_dq}) is fulfilled.

\begin{figure}[t]
\begin{center}
\includegraphics[width=13cm]{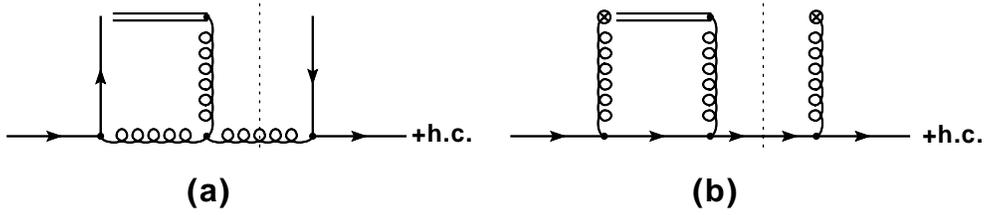}
\end{center}
\caption{One-loop diagrams relevant to the calculation of the Sivers function
for the quark (diagram (a)) and the gluon (diagram (b)) in the quark target 
model.
The symbol '$\otimes$' in diagram (b) indicates a Feynman rule for the field
strength tensor in~(\ref{e:gluon}) of the form 
$-i(p^\mu g^{\nu\rho} - p^{\nu}g^{\mu\rho}) \delta_{ab}$ where the indices
$\mu, \, \nu, \, a$ correspond to the indices of $F_a^{\mu\nu}$ and 
$\rho, \, b$ to those of the gluon in the figure 
(see also Ref.~\cite{collins_81c}).}
\label{f:qcd}
\end{figure}

In a second step we consider a quark target in perturbative QCD in order to 
study the QCD sum rule (\ref{e:rule}).
In the one-loop approximation we are using here only the quark and gluon
Sivers function are non-zero, while the antiquark Sivers function vanishes.
The relevant diagrams are depicted in Fig.~\ref{f:qcd}.
Although the algebra for the QCD calculation is more involved than in the
case of the spectator model the final results have a compact representation,
\begin{eqnarray} \label{e:res_q2}
f_{1T}^{\perp q}(x,\vec{k}_T^2) & = & - \frac{N_c C_F g^4}{4(2\pi)^4}
 \frac{x(1-x)m_q^2}{\vec{k}_T^2 (\vec{k}_T^2 + (1-x)^2 m_q^2)}  
 \, \ln \frac{\vec{k}_T^2 + (1-x)^2 m_q^2}{(1-x)^2 m_q^2} \,,
\\ \label{e:res_g2}
f_{1T}^{\perp g}(x,\vec{k}_T^2) & = & 
 - f_{1T}^{\perp q}(1-x,\vec{k}_T^2) \,,
 \vphantom{\frac{1}{1}}
\end{eqnarray}
with $C_F = (N_c^2 - 1)/(2N_c)$.
With exactly the same reasoning we used above for the diquark model one 
finds that the Burkardt sum rule~(\ref{e:rule}) is valid in this one-loop 
model.
Note that the sum rule is also fulfilled in QED for an electron target
in the one-loop approximation: as already mentioned the photon Sivers function
vanishes.
Moreover, the electron Sivers function is zero in this approximation, because 
a diagram like in Fig.~\ref{f:qcd}(a) does not exist in QED.
\\[0.3cm]
\noindent
{\bf 4.}~In summary, we have provided the first explicit check of the
Burkardt sum rule for the Sivers function.
According to this sum rule, inside a transversely polarized target 
the average transverse momentum of partons induced by the Sivers effect 
vanishes if one sums over all partons.
This sum rule may be considered to be as fundamental as the well-known
sum rule for the longitudinal momentum in the case of the forward 
unpolarized parton density.
For our check we have performed one-loop calculations of the Sivers 
function in two different models.
We have used a quark target in perturbative QCD and a scalar diquark
model for the nucleon. 
The Burkardt sum rule is fulfilled in both cases provided that one
regularizes the final integral upon the transverse parton momentum. 
\\[0.6cm]
\noindent
{\bf Acknowledgements:}
The work of S.M. has been supported by the German National Academic
Foundation, the work of M.S. by the Gra\-du\-ier\-ten\-kol\-leg
``Phy\-sik der Ele\-men\-tar\-teil\-chen an Be\-schleu\-ni\-gern und
im Uni\-ver\-sum.'' 
The work has also been partially supported by the
Ver\-bund\-for\-schung (BMBF) and the Trans\-re\-gio/SFB
Bo\-chum-Bonn-Gie\-ssen.
This research is part of the EU Integrated Infrastructure Initiative
Hadronphysics Project under contract number RII3-CT-2004-506078.


\end{document}